\documentclass[12pt]{article}
\usepackage{amsmath,epsfig,graphicx,amssymb}
\usepackage[margin=0.7in]{geometry}
\newcommand{\beq}{\begin{equation}}
\newcommand{\eeq}{\end{equation}}
\newcommand{\ben}{\begin{eqnarray}}
\newcommand{\een}{\end{eqnarray}}
\date{}
\begin{document}
\title{Possibility of Entanglement at LIGO!}
\author{Partha Ghose\footnote{partha.ghose@gmail.com} \\
The National Academy of Sciences, India,\\ 5 Lajpatrai Road, Allahabad 211002, India
\\and\\A. K. Rajagopal\\Inspire Institute Inc., Alexandria, VA, USA}
\maketitle
\begin{abstract}
It is shown that a linearized classical gravity wave $\hat{a}$ {\em la} Einstein can get entangled with an array of test masses in a plane perpendicular to its direction of propagation. A Bell-CHSH inequality based on the requirement of noncontextuality for classical realism is derived, and it is shown that the entangled state produced violates this inequality. 
\end{abstract}
\section{Introduction}
For weak gravitational fields Einstein's equation
\begin{equation}
R_{\mu\nu} = -8\pi G\left(T_{\mu\nu} - \frac{1}{2}g_{\mu\nu} T^\lambda_\lambda\right)
\end{equation}
can be linearized by writing the metric in the form
\begin{equation}
g_{\mu\nu}(x) = \eta_{\mu\nu} + h_{\mu\nu}(x)
\end{equation}
where $\eta_{\mu\nu}$ is the flat Minkowski metric and $h_{\mu\nu} \ll 1$ is a small perturbation. To the lowest order in the perturbation the vacuum equation $R_{\mu\nu} =0$ is then of the linearized form \cite{th, sch, gw}
\begin{equation}
\Box h_{\mu\nu} - \partial_\mu \xi_\nu - \partial_\nu \xi_\mu = 0
\end{equation}
where $\Box = \eta^{\alpha\beta}\partial_{\alpha\beta} = -\partial^2/\partial t^2 + \nabla^2$ and
\begin{equation}
\xi_\mu = \partial_\gamma h^\gamma_\mu - \frac{1}{2}\partial_\mu h^\gamma_\gamma. 
\end{equation}
This equation can be written in the form
\begin{eqnarray}
\Box h_{\mu\nu} = 0,\\
\partial_\nu h^\nu_\mu(x) - \frac{1}{2}\partial_\mu h^\nu_\nu = 0,\label{Lg}
\end{eqnarray}
where the first equation is the linearized Einstein equation and the second equation is the Lorentz gauge constraint, and where $h^\gamma_\nu \equiv \eta^{\gamma\delta} h_{\delta\nu}$. A general solution is of the form
\begin{equation}
h_{\mu\nu}(x) = \alpha_{\mu\nu}\,e^{ik_\lambda x^\lambda} + \alpha^*_{\mu\nu}\,e^{-ik_\lambda x^\lambda}
\end{equation}
with $k_\mu^\mu = 0$, $k^\mu = \eta^{\mu\nu}k_\nu$, where $\alpha_{\mu\nu}$ is the polarization tensor which is symmetric in $(\mu,\nu)$ and so has ten components, but this number can be reduced to two by making use of Bianchi identities and fixing the gauge. 

In the {\em transverse-traceless gauge} (TT gauge) the metric perturbation is made purely spatial by requiring $h_{tt} = h_{ti} = 0$ and also traceless, $h_i^{\,\,i} = 0$. Then the Lorentz gauge constraint (\ref{Lg}) becomes
\begin{equation}
\partial_i h_{ij} = 0,
\end{equation}
which ensures transversality. For propagation in the $z$ direction with a fixed frequency $\omega$ and $k^\mu = (\omega,0,0,\omega),\, k.x = \omega(z - t)$ and $c=1$, the general solution can be written as
\begin{equation}
h_{\mu\nu}(z) = \left(\begin{array}{cccc}
 0 & 0 & 0 & 0\\
0 &\alpha_{11} & \alpha_{12} & 0\\
0 & \alpha_{12} & -\alpha_{11} & 0\\
0 & 0 & 0 & 0
\end{array} \right) e^{i\omega(z - t)}\label{sol}
\end{equation}
which has two independent transverse polarization states and helicity $\pm 2$. The part proportional to $\alpha_{xx} = \alpha_{11}$ is called the {\em plus-polarization} state and is denoted by $+$, and the part proportional to $\alpha_{xy} = \alpha_{12} = \alpha_{21}$ is called the {\em cross-polarization} state and is denoted by $\times$. It is in this sense that one usually associates spin-2 with gravity.

It will be convenient for our purpose to write the above solution in the form
\begin{eqnarray}
h(z) &=& \left[f_\times(t - z)\varepsilon_\times + f_+ (t - z)\varepsilon_+\right]e^{i\omega(z - t)}\\
&\equiv& h_\times(z) + h_+(z) \label{hz}
\end{eqnarray}
where $h$ and $\epsilon$ are $4\times 4$ matrices with elements $h_{\mu\nu},\,\varepsilon_{\mu\nu}$ etc, and $h_\times(z),\, h_+(z)$ are respectively the purely cross-polarized and purely plus-polarized gravity waves along the $z$ axis, where 
\begin{equation}
\varepsilon_\times = \left(\begin{array}{cccc}
 0 & 0 & 0 & 0\\
0 & 0 & 1 & 0\\
0 & 1 & 0 & 0\\
0 & 0 & 0 & 0
\end{array} \right)
\end{equation}
\begin{equation}
\varepsilon_+ = \left(\begin{array}{cccc}
0 & 0 & 0 & 0\\
0 & 1 & 0 & 0\\
0 & 0 & -1 & 0\\
0 & 0 & 0 & 0
\end{array} \right)
\end{equation}
are the unit polarization tensors. Notice that $\epsilon_\times \epsilon_+ + \epsilon_+ \epsilon_\times = 0$. Hence, $h_\times(z)$ and $h_+(z)$ anti-commute. The physical implications of this property need to be investigated further.

Using the basis vectors $|+\rangle = \left(0,1,0,0\right)^T,\, |-\rangle = \left(0,0,1,0\right)^T$, one can construct $\varepsilon_\times = |+\rangle\otimes|-\rangle + |-\rangle\otimes|+\rangle \equiv |+ -\rangle + |- +\rangle$, and $\varepsilon_+ = |+\rangle\otimes|+\rangle - |-\rangle\otimes|-\rangle \equiv |+ +\rangle - |- -\rangle$. 
The other two basis vectors $\left(1,0,0,0\right)^T$ and $\left(0,0,0,1\right)^T$ are eliminated by choosing the TT gauge which makes the metric perturbation purely spatial and transverse. 

The inner product of two matrices $A$ and $B$ is defined by the Frobenius product
\begin{equation}
\langle A|B\rangle = \frac{1}{2}Tr (A^\dagger B).
\end{equation}
Hence, we have $\langle \varepsilon_+|\varepsilon_+\rangle = \langle \varepsilon_\times|\varepsilon_\times\rangle = 1,\,\langle \varepsilon_+|\varepsilon_\times\rangle = \langle \varepsilon_\times|\varepsilon_+\rangle = 0$.
\section{Schr\"{o}dinger Cat and Classical Gravity}
According to the Equivalence Principle the effects of gravity can be transformed away in a local enough region that tidal effects can be ignored. Therefore, the effect of gravity on a single test mass has no physical significance. But two or more test masses in different locations can be physically influenced by linearized gravity waves which can change their `proper distances'.  
Linearized gravity waves are transverse and can therefore change the proper distances between test masses in a plane perpendicular to their propagation direction.
If, therefore, a gravity wave like $h(z)$ (eqn (\ref{hz})) travelling in the $z$ direction is incident on an array of coplanar test masses at various locations in the $xy$ plane (as, for example, in each arm of a laser interferometer like LIGO), it can be shown \cite{th, gw} that the proper distances between the test masses will oscillate with the wave frequency $\nu = \omega/2\pi$ in two elliptical modes, one with axes parallel to the $(x,y)$ axes corresponding to the plus-polarization wave and the other with axes rotated by $\pi/4$ relative to the $(x,y)$ axes, corresponding to the cross-polarization wave. This $\pi/4$ rotation is because gravity is a rank-2 tensor field. Hence, the state of the total system (test masses + gravity wave) must be of the form
\begin{equation}
|H\rangle  = |h_+\rangle|(xy)_+\rangle + |h_\times\rangle|(xy)_\times\rangle \label{sch}
\end{equation}
where $|h_+\rangle$ and $|h_\times\rangle$ denote the two orthogonal polarization states of the gravity wave, and $|(xy)_+\rangle$ and $|(xy)_\times\rangle$ denote the corresponding oscillation modes of the two arms of the interferometer in the $xy$ plane. The $h_+$ component of the wave cannot produce $(xy)_\times$ oscillation modes and the $h_\times$ component cannot produce $(xy)_+$ oscillation modes, and therefore the states $|h_+\rangle|(xy)_\times\rangle$ and $|h_\times\rangle|(xy)_+\rangle$ cannot occur. Hence, (\ref{sch}) is a non-factorizable, and in that sense, `entangled' state of the two gravitational polarization states and their corresponding oscillation modes.  Since the interferometer arms are macroscopic, this is a Schr\"{o}dinger cat state before observation. The gravity waves span a Hilbert space $\mathcal{H}_g$ and the oscillatory interferometer states span a different Hilbert space $\mathcal{H}_{xy}$, and the state $H \in \mathcal{H}_g\otimes \mathcal{H}_{xy}$. 

The recent detection of gravitational waves may therefore be interpreted as indirect evidence of the production of entangled states in classical gravitational physics.

\section{Contextuality in Classical Gravity: A Bell-CHSH Test}
Let us consider a LIGO set up with the two arms of the laser interferometer in the $xy$ plane, and a gravitational wave incident on it along the $z$ direction. As we have seen from eqn (\ref{sol}), in the TT gauge there are two independent degrees of freedom and two amplitudes $\alpha_{xx} = -\alpha_{yy}$ and $\alpha_{xy} = \alpha_{yx}$. A wave with $\alpha_{xy} = 0$ produces a mteric
\begin{equation}
ds^2 = - dt^2 + (1 + h_+)dx^2 + (1 - h_+)dy^2 + dz^2
\end{equation}
where $h_+ = \alpha_{xx}\,{\rm exp}(i\omega(z - t))$. This produces opposite effects on the proper distance on the two axes, contracting one and expanding the other. On the other hand, if $\alpha_{xx} = 0$, only the off-diagonal terms $h_{xy} = h_{yx} = h_\times$ in the metric are non-zero, and that corresponds to a $\pi/4$ rotation relative to the previous case. A general wave is a linear superposition of these two, and depending on the phase relation, a circular or elliptical polarization is produced. Consequently, the proper distances between the test masses in the interferometer are stretched and compressed along the $x$ and $y$ directions periodically in two modes, one parallel to the $(xy)$ axes (the plus mode) and the other rotated by $\pi/4$ relative to the $(xy)$ axes (the cross mode). An interferometer in the $(xy)$ plane can measure the difference in the return times of light along the two arms in the $x$ and $y$ directions due to these changes in the proper distances through the phase changes they produce, resulting in fringe shifts in the interference pattern. Hence, the normalized state of the gravitational wave plus the oscillatory interferometer paths can be written in the form
\begin{eqnarray}
|\hat{H}\rangle &=& \frac{1}{\sqrt{2}} \left(|h_+\rangle|(xy)_+\rangle + |h_\times\rangle|(xy)_\times\rangle\right)\label{ent} 
\end{eqnarray}
where $|(xy)_+\rangle$ and $|(xy)_\times\rangle$ denote the two oscillation modes of the arms of the interferometer. 

To derive a Bell-CHSH inequality based on noncontextuality, let us first consider an arbitrary general state
\begin{eqnarray}
|H\rangle &=& (\cos\alpha|h_+\rangle+e^{i\beta}\sin\alpha|h_\times\rangle)(\cos\gamma|(xy)_+\rangle + e^{i\delta}\sin\gamma|(xy)_\times\rangle)\\
&\equiv& |\psi_h\rangle|\psi_{xy}\rangle\label{gen}
\end{eqnarray}
where $\alpha, \beta, \gamma, \delta$ are arbitrary parameters. Next, let us define the
correlation
\begin{eqnarray}
E(\theta,\phi)=\langle H\vert \sigma_{\theta}.\sigma_{\phi}\vert H\rangle
\end{eqnarray}
where $\vert H\rangle$ is an arbitrary normalized state of the apparatus + gravity, and
\begin{eqnarray}
\sigma_{\theta}=\sigma_{\theta,0}-\sigma_{\theta,\pi},\\
\sigma_{\phi}=\sigma_{\phi,0}-\sigma_{\phi,\pi},
\end{eqnarray}
with
\begin{eqnarray}
\sigma_{\theta,0}=\frac{1}{2}(|h_+\rangle + e^{i\theta}|h_\times\rangle)(\langle h_+| + e^{-i\theta}\langle h_\times|) \otimes \mathbb{I}_{xy},\nonumber \\
\sigma_{\theta,\pi}=\frac{1}{2}(|h_+\rangle - e^{i\theta}|h_\times\rangle)(\langle h_+| - e^{-i\theta} \langle h_\times|)\otimes \mathbb{I}_{xy},\nonumber \\
\sigma_{\phi,0}= \mathbb{I}_{h}\otimes\frac{1}{2}(|(xy)_+\rangle + e^{i\phi}|(xy)_\times\rangle)(\langle (xy)_+|+e^{-i\phi}\langle (xy)_\times|),\nonumber \\
\sigma_{\phi,\pi}= \mathbb{I}_{h}\otimes\frac{1}{2}(|(xy)_+\rangle-e^{i\phi}|(xy)_\times\rangle)(\langle (xy)_+|-e^{-i\phi}\langle (xy)_\times|),\label{sigma}
\end{eqnarray}
where $\theta$ and $\phi$ are phase shifts between the two polarization modes and the two path oscillation modes respectively, and $\mathbb{I}_{xy}$ and $\mathbb{I}_h$ are the identity operators in the Hilbert space $\mathcal{H}_{xy}$ spanned by the oscillating interferometer paths and the Hilbert space $\mathcal{H}_h$ spanned by the gravity wave polarizations respectively.
Hence,
\begin{eqnarray}
\sigma_{\theta}=(e^{-i\theta}|h_+\rangle\langle h_\times|+e^{i\theta}|h_\times\rangle \langle h_+|)\otimes \mathbb{I}_{xy},\\
\sigma_{\phi}= \mathbb{I}_{h}\otimes (e^{-i\phi}|(xy)_+\rangle\langle (xy)_\times|+e^{i\phi}|(xy)_\times\rangle\langle (xy)_+|).
\end{eqnarray} 
It should be noted that $\sigma_{\theta}$ and $\sigma_{\phi}$ act upon different Hilbert spaces altogether, and hence they commute with each other. This property is necessary for noncontextuality of the apparatus + gravity wave system. It has always been a tenet of classical physics that whatever exists in the physical world is independent of observations which only serve to reveal them. Put more technically, this means that the result of a measurement is predetermined and is not affected by how the value is measured, i.e. not affected by previous or simultaneous measurement of any other {\em compatible} or co-measureable observable. Hence the need for commuting observables to test noncontextuality.

For the general product state (\ref{gen}),
\begin{eqnarray}
E(\theta,\phi)&=&\langle\psi_h|\sigma_{\theta}|\psi_h\rangle\langle\psi_{xy}|\sigma_{\phi}|\psi_{xy}\rangle\nonumber \\
&=& E_h(\theta)E_{xy}(\phi),\label{exp}
\end{eqnarray}
with
\begin{eqnarray}
E_h(\theta)=\sin 2\alpha\cos(\beta-\theta),\label{epol}\\
E_{xy}(\phi)=\sin 2\gamma\cos(\delta-\phi).\label{epath}
\end{eqnarray}
Thus, the expectation value $E(\theta,\phi)$ is the product of the expectation values of the polarization and proper distance (or path) projections. Hence, the path and polarization measurements for product states in classical gravity are independent of one another in all contexts. This is the content of {\em noncontextuality}. This may, at first sight, look obvious and trivial, but on closer inspection, one finds that it implies the inequality
\beq
-1\leqslant E(\theta,\phi)\leqslant 1 \label{cor}
\eeq
for the correlation.

Now, define a quantity $S$ as
\ben
S(\theta_{1},\phi_{1};\theta_{2},\phi_{2}) &=& E(\theta_{1},\phi_{1})+E(\theta_{1},\phi_{2})-E(\theta_{2},\phi_{1})+E(\theta_{2},\phi_{2}).
\een
It follows from (\ref{cor}) that
\beq
|S|\leq 2.\label{bound}
\eeq
All that is required to derive this bound for product states is that {\em the correlations lie between $-1$ and $+1$}, which is guaranteed by the results (\ref{epol}) and (\ref{epath}). 
 
Now consider the correlation calculated for the normalized state (\ref{ent}), 
namely
\ben
E(\theta,\phi)&=&\langle \hat{H}|\sigma_{\theta}\cdot\sigma_{\phi}|\hat{H}\rangle\nonumber\\
&&\:= \langle \hat{H}\vert \,[(+)\sigma_{\theta,0} + (-)\sigma_{\theta,\pi} ]. [(+)\sigma_{\phi,0} + (-)\sigma_{\phi,\pi} ]\vert \hat{H}\rangle\\
&=&\langle \hat{H}[\sigma_{\theta,0}\cdot\sigma_{\phi,0}+\sigma_{\theta,\pi}\cdot\sigma_{\phi,\pi}\nonumber\\ &&\: -\sigma_{\theta,0}\cdot\sigma_{\phi,\pi}-\sigma_{\theta,\pi}\cdot\sigma_{\phi,0}]\vert \hat{H}\rangle
\een
The intensities corresponding to the four possible orientations are given by
\ben
I(\theta,\phi)&=& \langle \hat{H}|\sigma_{\theta,0}\cdot\sigma_{\phi,0}\vert \hat{H}\rangle,\nonumber\\
I(\theta+\pi,\phi+\pi)&=& \langle \hat{H}|\sigma_{\theta,\pi}\cdot\sigma_{\phi,\pi}\vert \hat{H}\rangle,\nonumber\\
I(\theta+\pi,\phi)&=& \langle \hat{H}|\sigma_{\theta,\pi}\cdot\sigma_{\phi,0}\vert \hat{H}\rangle,\nonumber\\
I(\theta,\phi+\pi)&=& \langle \hat{H}|\sigma_{\theta,0}\cdot\sigma_{\phi,\pi}\vert \hat{H}\rangle,\nonumber\\
\een
where clearly $I(\theta,\phi)=\frac{1}{2}[1+\cos(\theta+\phi)]$ from (\ref{ent}) and the definitions (\ref{sigma}). 

We can write $E(\theta, \phi)$ in terms of the normalized intensities as
\ben
E(\theta,\phi)&=&\dfrac{I(\theta,\phi)+I(\theta+\pi,\phi+\pi)-I(\theta+\pi,\phi)-I(\theta,\phi+\pi)}{I(\theta,\phi)+I(\theta+\pi,\phi+\pi)+I(\theta+\pi,\phi)+I(\theta,\phi+\pi)}\nonumber\\
&=& {\rm cos}(\theta + \phi).
\een
It is clear from this that the noncontextuality bound (\ref{bound}) is violated by the state (\ref{ent}) for the set of parameters $\theta_1 = 0, \theta_2 = \pi/2, \phi_1 = \pi/4, \phi_2 = - \pi/4$ for which $\vert S\vert = 2\sqrt{2}$. Since the distance and polarization changes occur in the same path, there is no violation of locality in this result. It is hoped that these predictions can one day be verified at facilities like LIGO.

Since we have used purely classical physics throughout, it is clear that {\em entanglement} and {\em contextuality} which are widely regarded as exclusively quantum mechanical, occur in classical gravity too. This is similar to entanglement and contextuality in classical polarization optics which is now a well established branch of optics \cite{spr, gh1, gh2, gh3, eb}.

\section{Acknowledgement}
The authors thank the S. N. Bose National Centre for Basic Sciences, Kolkata for inviting them to the {\em International Symposium on New Frontiers in Quantum Correlations}, Jan 29-Feb 2, 2018, where the basic ideas were developed. PG also thanks the National Academy of Sciences, India for a grant.


\begin{thebibliography}{0}
\bibitem{th}
Kip S. Thorne, `Gravitational Radiation: The Physics of Gravitational Waves and Their Generation', https:$//www.lorentz.leidenuniv.nl/lorentzchair/thorne/Thorne1.pdf$ 
\bibitem{sch}
B. F. Schutz and F. Ricci, `Gravitational Waves, theory and experiment (an overview)' in {\em Gravitational Waves} (2001) eds I. Ciufolini, V. Gorini, U. Moschella and P. Fr\'e, IoP Publishing, Bristol, 15-19. 
\bibitem{gw}
M. W. Guidry, Lectures on {\em Theoretical Astrophysics}, Chapter 21, http:
$//eagle.phys.utk.edu/guidry/astro616/lectures/lecture_ch21.pdf$
\bibitem{spr}
R. J. C. Spreeuw, {\em Found. of Phys.} {\bf 28} (1998) 361 . 
\bibitem{gh1}
P. Ghose and M. K. Samal, arXiv [quant-ph]/0111119 Nov 2001.
\bibitem{gh2}
P. Ghose and A. Mukherjee, {\em Rev. in Theoret. Sc.} {\bf 2} (2014) 1-14. 
\bibitem{gh3}
P. Ghose and A. Mukherjee, {\em Adv. Sc., Eng. and Med.} {\bf 6} (2014) 246-251.
\bibitem{eb}
J. H. Eberly et al, {\em Phys. Scripta} {\bf 91} (2016) 063003 and references therein.
\end{thebibliography}
\end{document}